\documentclass[10pt, preprintnumbers, aps, prd, twocolumn, superscriptaddress, nofootinbib]{revtex4-2}

\usepackage{array}
\usepackage{amssymb,amsmath,mathrsfs,enumerate}
\usepackage{bm}
\usepackage{graphicx}
\usepackage{feynmp-auto}
\usepackage{mathtools}
\usepackage{caption}
\usepackage{subcaption}
\usepackage[dvipsnames]{xcolor}
\usepackage[colorlinks=true, linkcolor=blue, urlcolor=Blue, citecolor=Mulberry]{hyperref}
\usepackage{orcidlink}
\usepackage{multirow}
\usepackage{booktabs}
\usepackage[normalem]{ulem}
\usepackage{microtype}  % nicer spacing (optional)
\usepackage{ragged2e}   % for \plotalign

\newcommand{\blfootnote}[1]{%
  \begingroup
  \renewcommand{\thefootnote}{}\footnote{#1}%
  \addtocounter{footnote}{-1}%
  \endgroup
}

\newcommand{\dashedline}{%
  \noalign{\vskip 2pt}%
  \multispan{4}\leaders\hbox{\rule{2pt}{0.3pt}\kern2pt}\hfill\cr
  \noalign{\vskip 2pt}%
}

\makeatletter
\patchcmd{\@makecaption}{\ignorespaces}{\justifying\ignorespaces}{}{}
\makeatother

\begin{document}

\title{LUX-ZEPLIN's Stairway to Hea$\nu$en: \\
 limits on elastic scatters of dark matter from solar capture}

\author{Debajit Bose \orcidlink{0000-0001-8594-8885}}
\email{debajitbose550@gmail.com}
\affiliation{Centre for High Energy Physics, Indian Institute of Science, C. V. Raman Avenue, Bengaluru 560012, India}
\author{Thong T.Q. Nguyen \orcidlink{0000-0002-8460-0219}}
\email{thong.nguyen@fysik.su.se}
\affiliation{Stockholm University and The Oskar Klein Centre for Cosmoparticle Physics, Alba Nova, 10691 Stockholm, Sweden}
\author{Akash Kumar Saha \orcidlink{0000-0002-3033-2589}}
\email{akashks@iisc.ac.in}
\affiliation{Centre for High Energy Physics, Indian Institute of Science, C. V. Raman Avenue, Bengaluru 560012, India}
\affiliation{Raymond and Beverly Sackler School of Physics and Astronomy, Tel Aviv University, Tel-Aviv 69978, Israel}
\affiliation{Department of Physics and Astronomy, University of California - Irvine, Irvine, CA 92697, U.S.A.}
\author{Nirmal Raj \orcidlink{0000-0002-4378-1201}}
\email{nraj@iisc.ac.in}
\affiliation{Centre for High Energy Physics, Indian Institute of Science, C. V. Raman Avenue, Bengaluru 560012, India}
\author{Tarak Nath Maity \orcidlink{0000-0003-1069-6358}}
\email{tarak.maity.physics@gmail.com}
\affiliation{Theory Division, Saha Institute of Nuclear Physics, 1/AF, Bidhannagar, Kolkata 700064, India}
\affiliation{Homi Bhabha National Institute, Training School Complex, Anushaktinagar, Mumbai-400094, India}
\author{Ranjan Laha \orcidlink{0000-0001-7104-5730}}
\email{ranjanlaha@iisc.ac.in}
\affiliation{Centre for High Energy Physics, Indian Institute of Science, C. V. Raman Avenue, Bengaluru 560012, India}
\date{\today}
%
%
%==================% ABSTRACT ==================%
\begin{abstract}
The recent report of xenon recoiling with about $250\, {\rm keV}$ energy at the LUX-ZEPLIN (LZ) dark matter direct detection experiment was interpreted by the collaboration as either a momentum-dependent elastic scatter or an inelastic scatter. 
In this study, we take elastic operators listed by LZ that fit the datum to better than 3$\sigma$, and perform a log-likelihood analysis using the reported significances to reverse-engineer their best-fit Wilson coefficients. 
We then estimate for each operator the rate of dark matter capturing in the Sun, and assuming self-annihilations of the captured
population to various Standard Model final states studied by IceCube, we place limits on them from measurements at IceCube and Super-Kamiokande of the fluxes of solar-direction and atmospheric neutrinos.
We show that large ranges of dark matter masses around the weak scale that explain the LZ event are ruled out in this scenario, with the strongest limits placed by direct annihilations to neutrino-antineutrino final states.
\end{abstract}
%===============================================%

\maketitle
\blfootnote{D.B., T.T.Q.N. and A.K.S. contributed equally to this work.}

%=================== INTRO =====================%
\section{Introduction}
\label{sec:intro}

With 2.84~tonne-year exposure and an extension of its search window to 270 keV nuclear recoils, LUX-ZEPLIN (LZ), a dual-phase xenon dark matter (DM) direct detection experiment, has reported an unexplained event, with a nuclear recoil energy of 248 $\pm$ 23 (stat) $\pm$ 23 (sys) keV~\cite{LZ:2026axp}. 
The event was recorded on June 16, 2023, and hence we denote it as LZ230616.
Its significance after accounting for look-elsewhere effects is 2.6$\sigma$. 
LZ also reported statistical fits of the event to various effective operators for DM-nucleon interactions taken from Refs.~\cite{Anand:2013yka,Barello:2014uda}: in Table~\ref{tab:opbenchmarks} we collect those with (local) significance $\geq3\sigma$. 
Elastic scattering via a Lorentz-invariant structure ($\mathcal{L}_{10}$) that arises from a magnetic dipole interaction of DM with the photon, giving rise to a momentum-dependent spin-spin coupling in the non-relativistic limit, 
and inelastic scattering via Galilean-invariant scalar-scalar ($\mathcal{O}_1$) and spin-spin ($\mathcal{O}_4$) operators, are the most favored at 3.4$\sigma$.
%
%
%=================== Fig: 1 =====================%
\begin{figure}[t]
    \centering
    \includegraphics[width=0.975\linewidth]{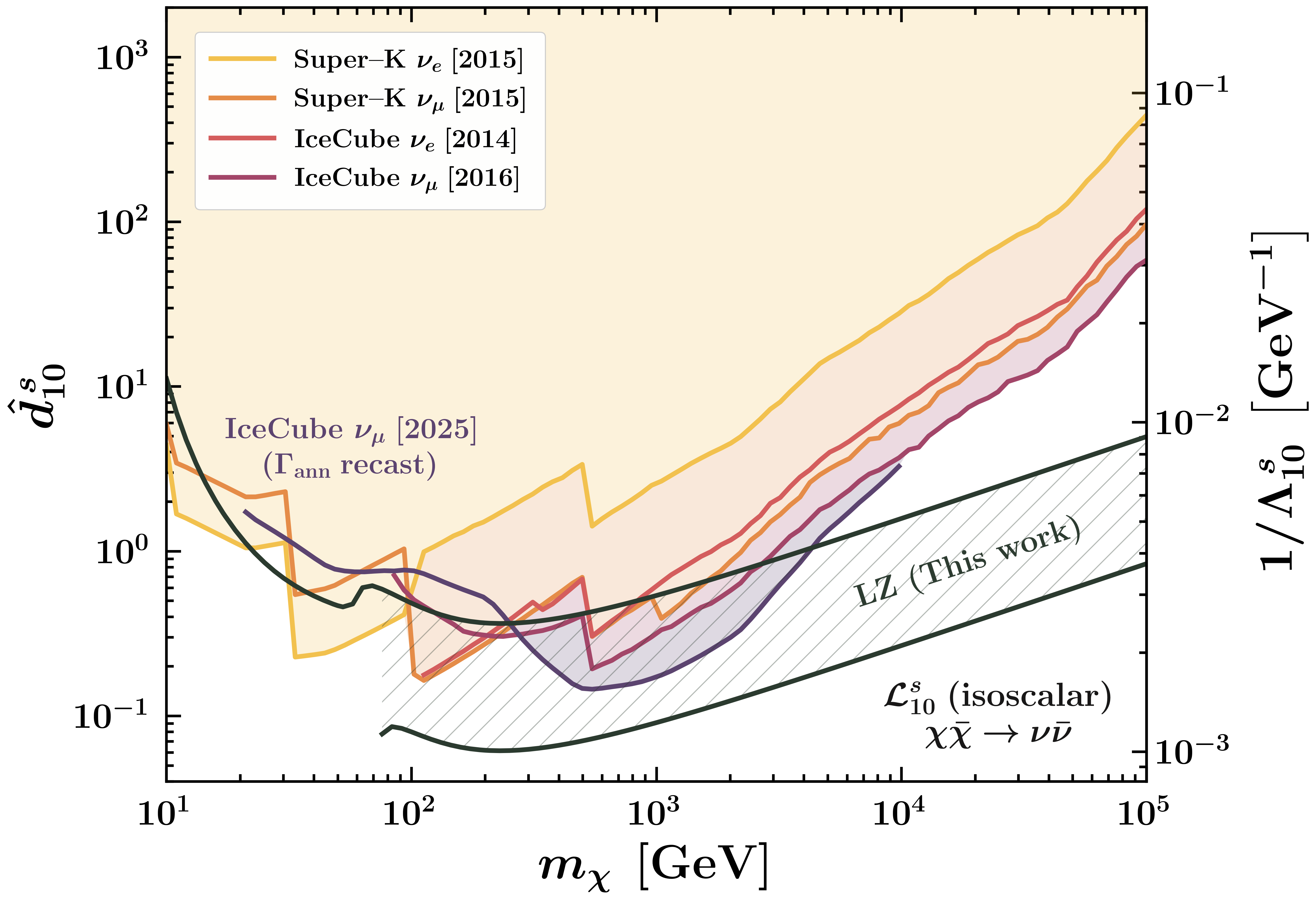}
    \caption{\justifying 95\% C.L. limits from IceCube and Super-Kamiokande (Super-K) on solar capture of dark matter via the operator $\mathcal{L}_{10}^s$ followed by annihilations entirely to $\nu\bar\nu$.
    These limits rule out part of the parametric regions favored by the hard-recoil event LZ230616.
    In the rest of the paper we obtain best-fits of LZ230616 to operators listed in Table~\ref{tab:opbenchmarks}, derive solar capture limits for other DM annihilation final states, and come to similar conclusions for most operators.
    See text for further details.}
    \label{fig:Ls10_capture_limits_nunu}
\end{figure}
%================================================%
%
%

Generally, elastic momentum-independent scatters are in tension with the LZ dataset as they would have produced numerous events at lower recoils than LZ230616~\cite{LZ:2026axp, DiMauro:2026ldr}.
In this study, we constrain the benchmarks in Table~\ref{tab:opbenchmarks} from capture of DM in the Sun facilitated by {\em elastic} operators; this was done for inelastic operators in Ref.~\cite{Nguyen:2026lui}.
One major undertaking of our work is to use the fit significances provided by LZ in Ref.~\cite{LZ:2026axp} as inputs to reconstruct our own best-fit parameters for all elastic operators using a log-likelihood analysis; such best-fit parameters were shown by LZ only for the operators $\mathcal{L}_{10}^s$ and $\mathcal{O}_1^s$.
This non-relativistic effective operator formalism for direct detection is motivated by UV completions involving models of weakly interacting massive particles (WIMPs), which are generically expected to annihilate to Standard Model (SM) states.
Therefore, post-capture, we assume that DM annihilates to one set of final states from the commonly studied $\{W^+W^-,b\bar{b},\tau^+\tau^-,\nu\bar{\nu}\}$ as constrained by IceCube~\cite{IceCube:2021xzo, IceCube:2025fcu}, and as assumed in Ref.~\cite{Nguyen:2026lui}.
We then compare the flux of neutrinos produced in these final states with measurements of solar-direction neutrinos at IceCube~\cite{IceCube:2016dgk,IceCube:2025fcu}, as well as atmospheric neutrinos at IceCube and Super-K~\cite{IceCube:2015mgt, Super-Kamiokande:2015qek}.
We thereby constrain large regions of parameter space for various operators, motivating further scrutiny of the unconstrained windows.
In Fig.~\ref{fig:Ls10_capture_limits_nunu} we illustrate this with our result for the operator $\mathcal{L}_{10}^s$ with the $\nu\bar{\nu}$ final state, which we will elaborate on in Sec.~\ref{sec:analysisresults}. Throughout this work, $\nu_\alpha$ represents the combined neutrino and antineutrino contributions of flavor $\alpha$, since Super-K and IceCube do not distinguish between them at the energies considered.

Solar capture and annihilations were used to strongly constrain the reading of LZ230616 as an inelastic scatter of the 1.1~TeV thermal Higgsino sourced by the Standard Halo Model in Refs.~\cite{Pospelov:2026ewn, Bose:2026ndd, DiMauro:2026dqp, DiMauro:2026ymt, Capolupo:2026uyi, Girmohanta:2026qzo, Murayama:2026apt}, though room was still left open for non-thermal Higgsinos~\cite{Bose:2026ndd} and/or non-standard halo velocities~\cite{Fan:2026kxx, DiMauro:2026ldr}.
These conclusions applied also to other inelastic scattering scenarios constrained in Ref.~\cite{Catena:2018vzc, Nguyen:2026lui}, to which we refer the reader for further discussion on how limits of this sort may be evaded. 
LZ230616 has naturally attracted an array of interesting interpretations~\cite{Fan:2026kxx, Freese:2026sga, Wu:2026nhi, Yin:2026jnn, DiMauro:2026ldr, Du:2026guj,Visinelli:2026kgt, Yin:2026jnn, Jeesun:2026vzo, Gu:2026vto, Smirnov:2026aqk, Unwin:2026rdp, Lou:2026idn, Su:2026rwz, Yamashita:2026ump, McCabe:2026crm, Chattopadhyay:2026ryw, deLima:2026shq, Baer:2026fpy, Lee:2026wof, Wang:2026ytg, Yang:2026wpb, Das:2026uyy, Alhazmi:2026efz, Okada:2026eol, Ahmed:2026qjg, Du:2026lpa, Bandyopadhyay:2026gjw, Kannike:2026qyl, Borah:2026zwf, Bisal:2026khf, 2026arXiv260908712C, 2026arXiv260908893Y, 2026arXiv260908993E, 2026arXiv260909015Z, 2026arXiv260909107A, 2026arXiv260909136L, 2026arXiv260909138L, Langhoff:2026ujr, Chatterjee:2026scv, He:2026hqz, Fan:2026hzw, Qi:2026vyp, Frolovsky:2026tvq, Kumar:2026lgi, Nagata:2026pbj, An:2026pkc, Arcadi:2026kev, Xing:2026civ, Uttayarat:2026isp, Baer:2026yrt, Lian:2026hpm, He:2026idw, Palmisano:2026kuj, Mahapatra:2026glu, Das:2026buc, Le-Yaouanc:2026djt, Ghosh:2026txe, Barman:2026omh, Borah:2026ris, Cabo-Almeida:2026uqw, Okada:2026upm, Lee:2026zbr, Heikinheimo:2026kwp, Kotlarski:2026pep, Okada:2026fef, Paul:2026okh, Ahmed:2026com, De:2026win, Khan:2026zuj, Sheng:2026tqt, Chauhan:2026udz, Sannino:2026hkc, Khan:2026osp, Jung:2026otm, Maier:2026qpr, Tang:2026lgu, Nomura:2026qyq, Yuan:2026djt, Cheung:2026byg, Elahi:2026vlm, Zhu:2026dag, Aghaie:2026vsu, Lee:2026jxl, Asadi:2026iot, Lee:2026xxh, Ahmed:2026kan, Gemmell:2026yaw, Delepine:2026ith, Langhoff:2026qgo, Brdar:2026ukx, Kalita:2026hvb, Chanda:2026wdl, Paul:2026okh, Roy:2026wzb}.
Meanwhile, observations of neutrinos from the Sun provide a powerful avenue for constraining DM interactions~\cite{Tucker-Smith:2001myb, Nussinov:2009ft, Menon:2009qj, IceCube:2016dgk, IceCube:2021xzo, Maity:2023rez, Chauhan:2023zuf, Bose:2023yll, Krishna:2025ncv, Nguyen:2025ygc, IceCube:2025fcu, Nguyen:2026nhe, Nguyen:2026apa, Bose:2026yun, Nguyen:2026pdr}. 
%
%================ END: INTRO ===================%
%

%
%
%===================== Table: 1  journal ========================%
\begin{table*}[htbp]
\centering
\begin{tabular}{cccc}
\hline\hline
elastic & operator & max& $m_\chi$/GeV [$\geq 3\sigma$] \\
operator~\cite{Anand:2013yka} & label & local $\sigma$  & \\
\hline
\noalign{\vskip 2pt}
\multirow{2}{*}{$(\bar\chi\, \chi)\,(\bar N\,i\gamma^5 N)$} & $\mathcal{L}^{s}_{2}$  & 3.1 & 1000, 4000 \\
& $\mathcal{L}^{v}_{2}$  & 3.1 & 400, 1000, 4000 \\
\dashedline
\multirow{2}{*}{$(\bar\chi\, \gamma^5 \chi)\,(\bar N\, \gamma^5 N)$} & $\mathcal{L}^{s}_{4}$  & 3.1 & 200, 400, 1000, 4000 \\
& $\mathcal{L}^{v}_{4}$  & 3.2 & 200, 400, 1000, 4000 \\
\dashedline
$(\bar\chi\,\gamma^\mu\,\chi)\,(\bar N i \sigma_{\mu\nu} \frac{q^\nu}{m_N} N)$ & $\mathcal{L}^{v}_{6}$  & 3.3 & 200, 400, 1000, 4000 \\
\dashedline
\multirow{2}{*}{$(\bar\chi\,\gamma^\mu\,\chi)\,(\bar N\,i \sigma_{\mu \nu} \tfrac{q_\nu}{m_N} \gamma^5 N)$} & $\mathcal{L}^{s}_{8}$  & 3.0 & 1000, 4000 \\
& $\mathcal{L}^{v}_{8}$  & 3.0 & 1000, 4000 \\
\dashedline
\multirow{2}{*}{$\bigl(\bar\chi\,i\sigma^{\mu\nu}\tfrac{q_\nu}{m_N}\chi\bigr)
  (\bar N\, \gamma_\mu \,N)$ } & $\mathcal{L}^{s}_{9}$  & 3.1 & 1000, 4000 \\
& $\mathcal{L}^{v}_{9}$  & 3.2 & 200, 400, 1000, 4000 \\
\dashedline
\multirow{2}{*}{$\bigl(\bar\chi\,i\sigma^{\mu\alpha}\tfrac{q_\alpha}{m_N}\chi\bigr)
  \bigl(\bar N\,i\sigma_{\mu\beta}\tfrac{q^\beta}{m_N} N\bigr)$ } & $\mathcal{L}^{s}_{10}$ & 3.4 & 200, 400, 1000, 4000 \\
& $\mathcal{L}^{v}_{10}$ & 3.4 & 100, 200, 400, 1000, 4000 \\
\dashedline
\multirow{2}{*}{$\bigl(\bar\chi\,i\sigma^{\mu\nu}\tfrac{q_\nu}{m_N}\chi\bigr)
   (\bar N\,\gamma_\mu\gamma^5 N)$ }& $\mathcal{L}^{s}_{11}$ & 3.2 & 200, 400, 1000, 4000 \\
& $\mathcal{L}^{v}_{11}$ & 3.2 & 400, 1000, 4000 \\
\dashedline
\multirow{2}{*}{$\bigl(i \bar\chi\,i\sigma^{\mu\nu}\tfrac{q_\nu}{m_N}\chi\bigr)
  (\bar N\,i \sigma_{\mu \alpha} \tfrac{q^\alpha}{m_N} \gamma^5 N)$} & $\mathcal{L}^{s}_{12}$ & 3.3 & 200, 400, 1000, 4000 \\
& $\mathcal{L}^{v}_{12}$ & 3.2 & 200, 400, 1000, 4000 \\
\dashedline
 $(\bar\chi\,\gamma^\mu\gamma^5\chi)\, 
   (\bar N \gamma_\mu N)$ & $\mathcal{L}^{v}_{13}$ & 3.0 & 400, 1000, 4000 \\
\dashedline
\multirow{2}{*}{$(\bar\chi\,\gamma^\mu\gamma^5\chi)\,
   \bigl(\bar N\,i\sigma_{\mu\nu}\tfrac{q^\nu}{m_N} N\bigr)$} & $\mathcal{L}^{s}_{14}$ & 3.1 & 400, 1000, 4000 \\
& $\mathcal{L}^{v}_{14}$ & 3.1 & 400, 1000, 4000 \\
\dashedline
\multirow{2}{*}{$(i \bar\chi\,\gamma^\mu\gamma^5\chi)\,
  (\bar N\,i \sigma_{\mu \nu} \tfrac{q^\nu}{m_N} \gamma^5 N)$} & $\mathcal{L}^{s}_{16}$ & 3.4 & 200, 400, 1000, 4000 \\
& $\mathcal{L}^{v}_{16}$ & 3.3 & 200, 400, 1000, 4000 \\
\dashedline
$(i\bar{\chi}i\sigma^{\mu \nu} \tfrac{q_\nu}{m_N}\gamma^5\chi)\bigl(\bar N\,i \sigma_{\mu \alpha} \tfrac{q^\alpha}{m_N} N \bigr)$ & $\mathcal{L}^{s}_{18}$ & 3.1 & 400, 1000, 4000 \\
\dashedline
\multirow{2}{*}{$(i\bar{\chi}i\sigma^{\mu \nu} \tfrac{q_\nu}{m_N}\gamma^5\chi)
   \bigl(\bar N\,\gamma_\mu\gamma^5 N\bigr)$ } & $\mathcal{L}^{s}_{19}$ & 3.1 & 400, 1000, 4000 \\
& $\mathcal{L}^{v}_{19}$ & 3.1 & 400, 1000, 4000 \\
\dashedline
\multirow{2}{*}{$(i\bar{\chi}i\sigma^{\mu \nu} \tfrac{q_\nu}{m_N}\gamma^5\chi)\bigl(\bar N\,i \sigma_{\mu \alpha} \tfrac{q^\alpha}{m_N} \gamma^5 N \bigr)$ } & $\mathcal{L}^{s}_{20}$ & 3.3 & 200, 400, 1000, 4000 \\
& $\mathcal{L}^{v}_{20}$ & 3.3 & 200, 400, 1000, 4000 \\
\noalign{\vskip 2pt}
\hline
\hline
inelastic & operator & max & $m_\chi$/GeV$_{\delta/{\rm keV}}$ [$\geq 3\sigma$] \\
operator~\cite{Barello:2014uda} & label & local $\sigma$ & \\
\hline
\noalign{\vskip 2pt}
\multirow{2}{*}{$\mathbf{1}_\chi\,\mathbf{1}_N$} & $\mathcal{O}^{s}_{1}$ & 3.3 & 1000$_{300-350}$,4000$_{300-350}$  \\
& $\mathcal{O}^{v}_{1}$ & 3.4 & 400$_{250-350}$, 1000$_{250-350}$, 4000$_{250-350}$  \\
\dashedline
\multirow{2}{*}{$\vec{S}_\chi\!\cdot\!\vec{S}_N$} & $\mathcal{O}^{s}_{4}$ & 3.4 & 400$_{150-300}$, 1000$_{150-350}$, 4000$_{50-350}$ \\
& $\mathcal{O}^{v}_{4}$ & 3.4 & 400$_{150-300}$, 1000$_{150-350}$, 4000$_{50-350}$  \\
\noalign{\vskip 2pt}
\hline\hline
\end{tabular}
\caption{\justifying {\bf \em Top}: NREFT elastic operators reaching a local significance
$\geq 3\sigma$ as per LZ~\cite{LZ:2026axp}. 
We quote here only the maximum significance across DM masses specified by LZ, with the last column listing masses at which the operator clears $3\sigma$.
{\bf \em Bottom}: Same, but with NREFT inelastic operators.
Along with the DM mass, we also specify the range of inter-state mass splittings $\delta$ clearing $3\sigma$.
Here $s$ and $v$ denote respectively the isoscalar and isovector projection of the operator.}
\label{tab:opbenchmarks}
\end{table*}
%=================== END: Table: 1 journal =====================%

%
%
%=================== Fig: 2 =====================%
\begin{figure*}[t]
    \centering
   \begin{subfigure}{0.495\textwidth}
        \centering
        \includegraphics[width=0.975\linewidth]{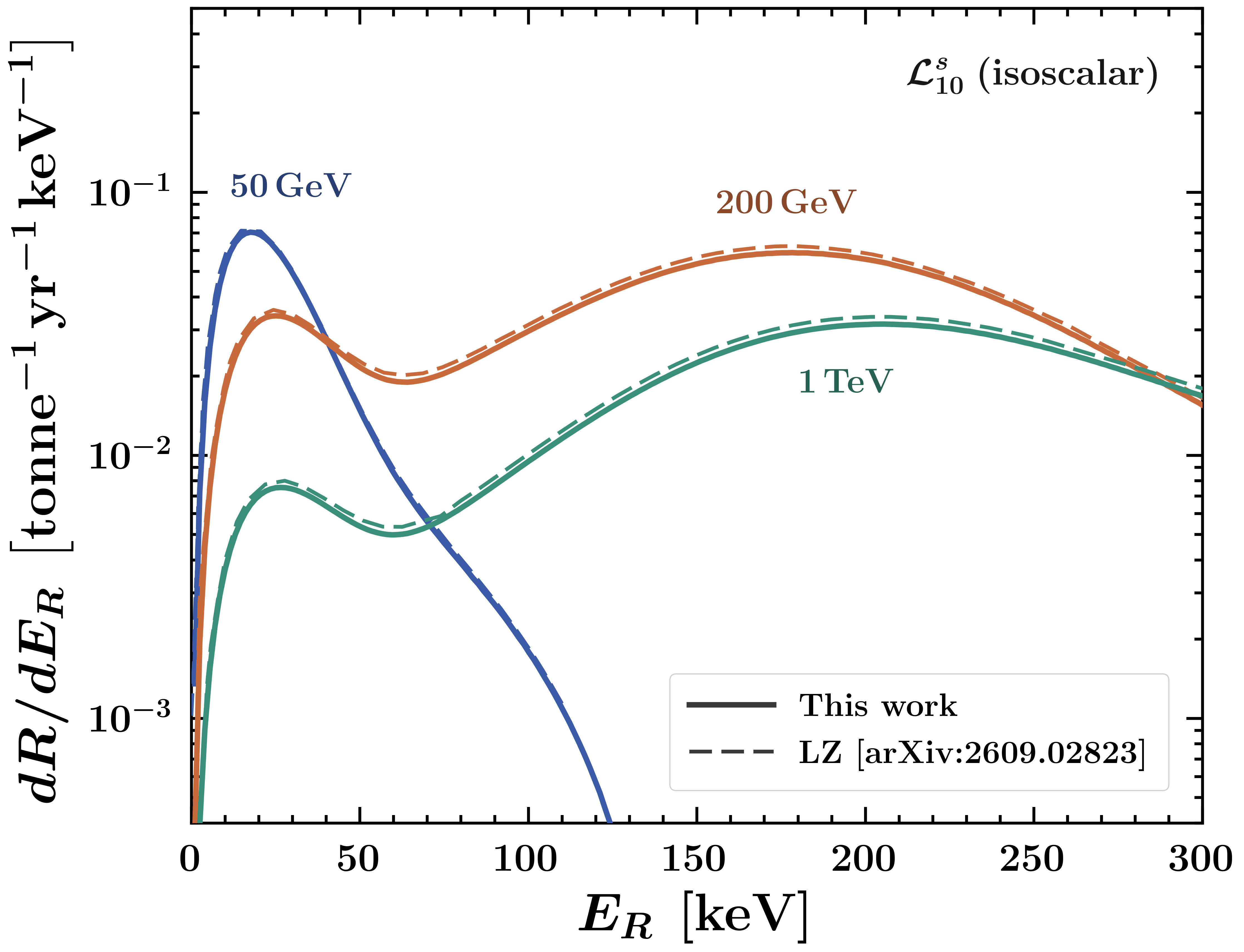}
        \label{sf:Ls10_recoil_comparison_LZ}
        % \caption{}
     \end{subfigure}
    \begin{subfigure}{0.495\textwidth}
        \centering
        \includegraphics[width=0.975\linewidth]{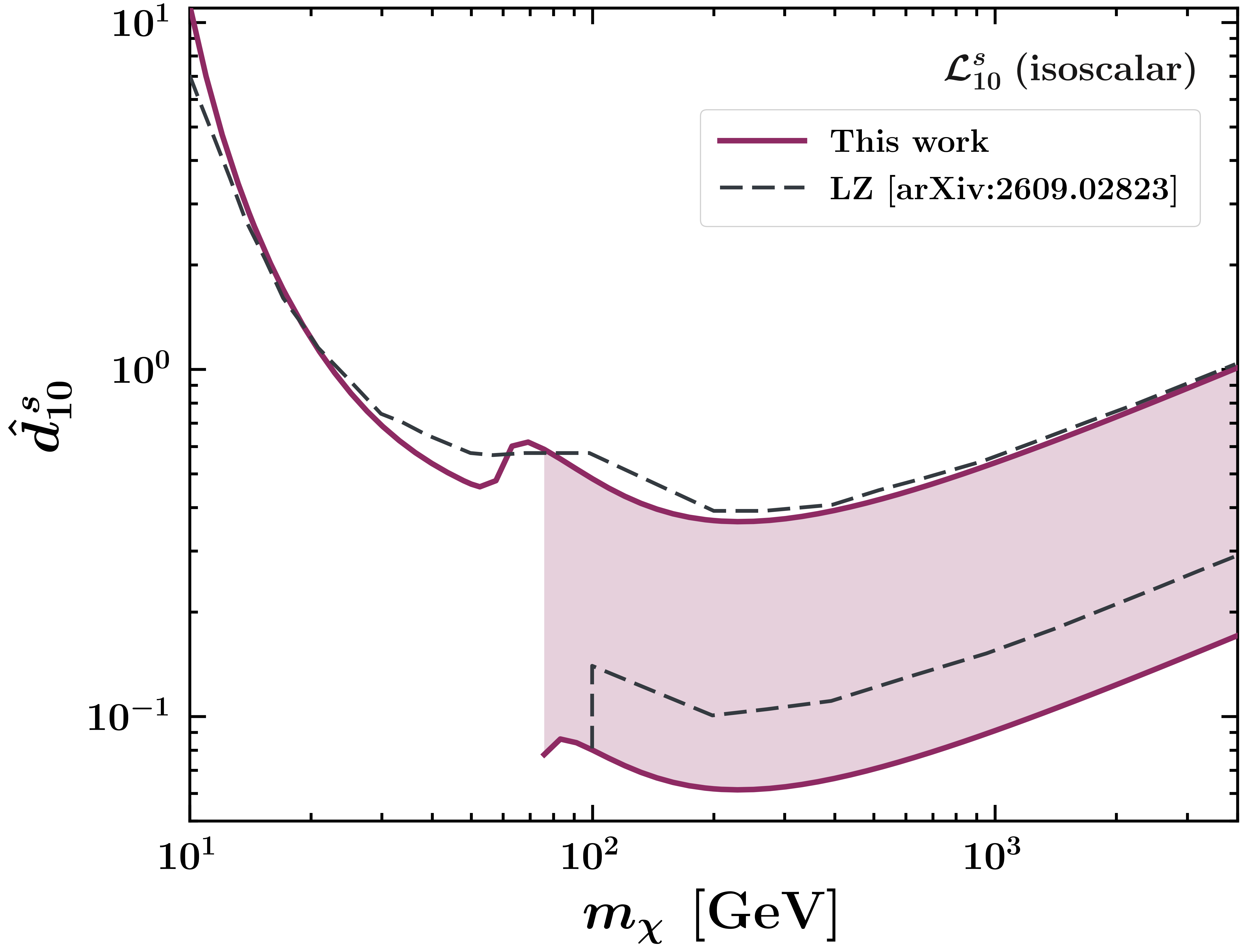}
        \label{sf:Ls10_limit_comparison_LZ}
        % \caption{}
    \end{subfigure}
\par\vspace{0.4cm}
    \begin{subfigure}{0.495\textwidth}
        \centering
        \includegraphics[width=0.975\linewidth]{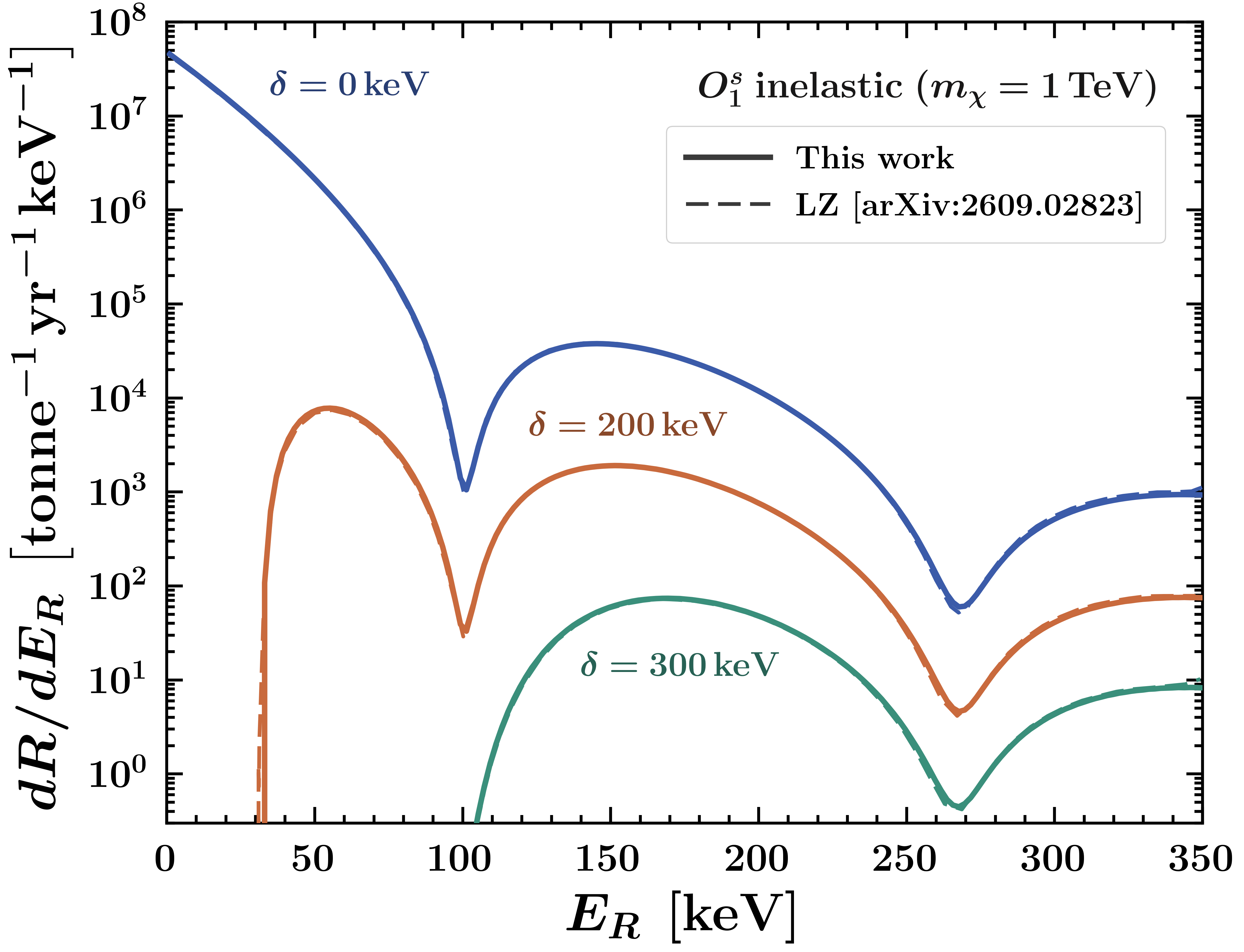}
        \label{sf:O1s_inelastic_recoil_comparison_LZ}
        % \caption{}
    \end{subfigure}
    \begin{subfigure}{0.495\textwidth}
        \centering
        \includegraphics[width=0.975\linewidth]{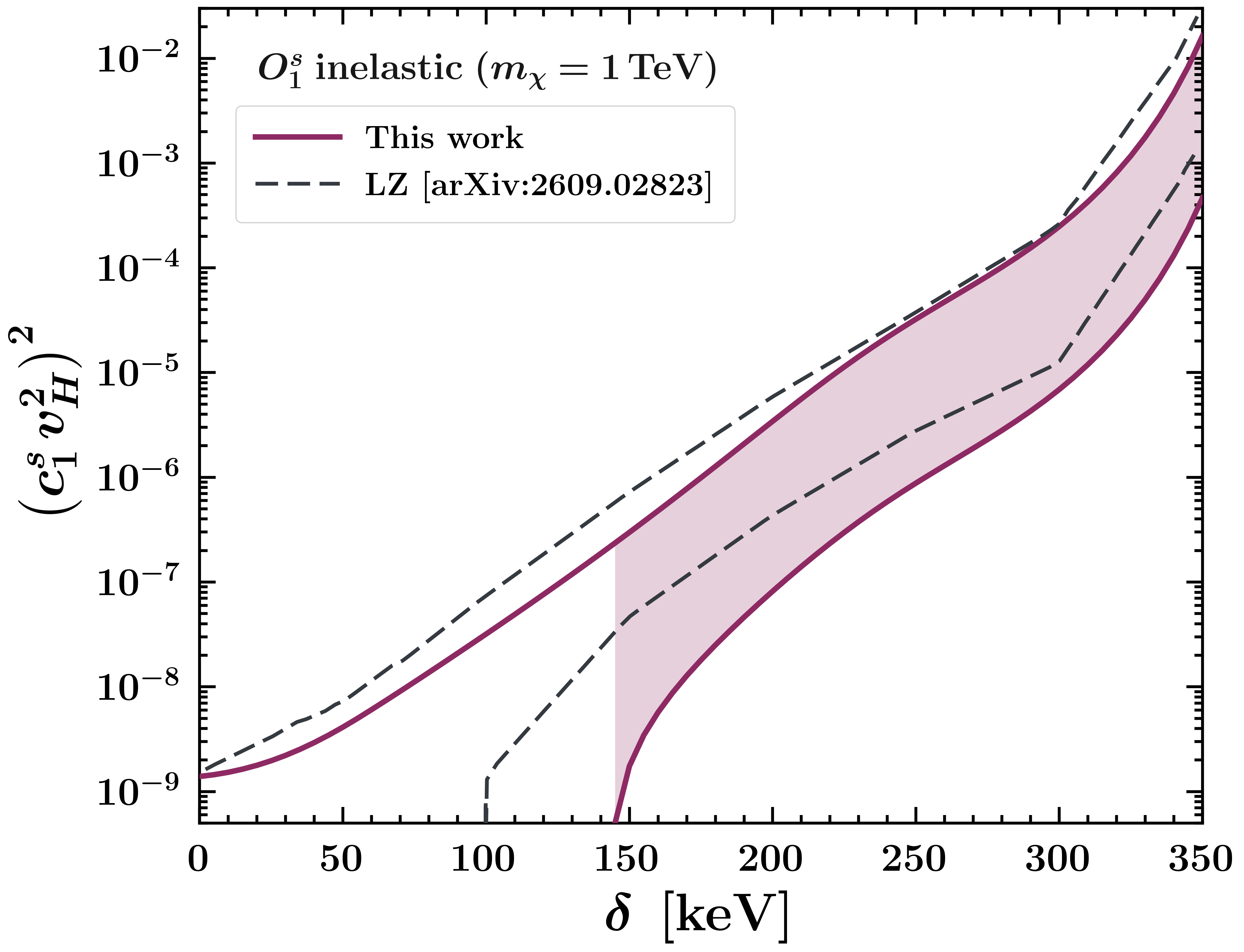}
        \label{sf:O1s_inelastic_limit_comparison_LZ}
        % \caption{}
    \end{subfigure}
%
%%%%%
    \caption{\justifying {\bf \em Left}: Recoil energy spectra for the effective operators that were fit by the LZ collaboration at 3.4$\sigma$ significance to the LZ230616 event and displayed in Ref.~\cite{LZ:2026axp}. 
    The top panel shows elastic scattering via the momentum-dependent spin-spin operator that may arise from a DM dipole moment and the bottom panel inelastic scattering via the scalar-scalar operator, with the isoscalar projection taken for both. 
    The spectra we obtain are seen to agree well with those provided by LZ.
    {\bf \em Right}: Best-fit parameters and 90\% confidence intervals for the operators corresponding to the left panel, computed using the log-likelihood procedure outlined in Sec.~\ref{sec:lzlikelihoods} taking LZ's reported fit significances as the input.
    The factor-of-few agreement of our fits with LZ gives us license to use our procedure to obtain best-fit parameters for all other elastic operators considered in this work, listed in Table~\ref{tab:opbenchmarks}.
    Our limits from solar capture may then be compared to our fits.
    See Sec.~\ref{sec:lzlikelihoods} for further details.}
    \label{fig:L10_O1s_dd_limits_LZ}
\end{figure*}
%================================================%
%
%

%================ Direct detection ===================%
\section{Reconstructing the LZ fits}
\label{sec:lzlikelihoods}

In this section we specify the recipe by which we identify regions in the space of elastic operator Wilson co-efficients $\{\hat{d}\}$ vs DM mass $m_\chi$ that provide fits to LZ230616 at the significance reported by LZ in Ref~\cite{LZ:2026axp}.
Such regions were displayed by LZ only for the magnetic moment-sourced isoscalar operator $\mathcal{L}^s_{10}$ and the inelastic scalar-scalar isoscalar operator $\mathcal{O}_1^s$, which gave 3.4$\sigma$ fits~\cite{LZ:2026axp}.
Thus we use our recipe to first reconstruct the favored region for $\mathcal{L}_{10}^s$, finding good agreement with LZ's displayed region.
We then repeat our method for all other operators of interest.

First, we compute the nuclear recoil spectra using
\texttt{WimPyDD} package~\cite{Jeong:2021wimpydd}, which implements
DM--nucleus scattering in the Galilean-invariant non-relativistic
effective field theory (NREFT) framework~\cite{Fitzpatrick:2012ix,Anand:2013yka,Kang:2018odb}. The interaction Hamiltonian is~\cite{Gondolo:2020wge, Jeong:2021wimpydd}
\begin{equation}\label{eq:nreft_hamiltonian}
    \mathcal{H}(\mathbf{r}) = \sum_{\tau=0,1}\sum_i c_i^\tau(m_\chi,q)\, \mathcal{O}_i(\mathbf{r})\,t^\tau,
\end{equation}
where $q$ is the momentum transfer, $\mathcal{O}_i$ are the NREFT operators,
$t^0=1$, and $t^1=\tau_3$, with $\tau_3=\mathrm{diag}(1,-1)$.
The Hamiltonian is a diagonal matrix, with its two entries corresponding to the proton and neutron couplings, $c_i^p=(c_i^0+c_i^1)/2$ and $c_i^n=(c_i^0-c_i^1)/2$, respectively.
For each relativistic interaction $\mathcal{L}j^{s,v}$, where $s$ and $v$ denote isoscalar and isovector interactions, respectively, we retain all NREFT operators, momentum and $m_\chi$ dependences, and interference terms generated by its non-relativistic reduction.
The mappings $\mathcal{L}_j\rightarrow\sum_i c_{i,j}\mathcal{O}_i$ are taken from Table~I of the journal version of Ref.~\cite{Anand:2013yka}, with $m_M=m_N$.\footnote{The journal and last-posted pre-print versions of Ref.~\cite{Anand:2013yka} differ in the covariant structure of several operators and their non-relativistic reductions. 
Throughout this work, we use the journal-version mappings as adopted by LZ.}
The dimensionless couplings $\widehat d_j^{s,v}$ associated with the Lagrangians $\mathcal{L}_j^{s,v}$ are defined as
\begin{equation}
    \widehat d_j^{s,v} \equiv v_{\rm  H}^2 d_j^{s,v} \, ,
    % \qquad
    % v_{\rm H}=246.2~\mathrm{GeV}.
    \label{eq:lagrangian_normalization}
\end{equation}
where $v_{\rm H}=246.2~\mathrm{GeV}$ is the Higgs vacuum expectation value. This normalization applies to the operators involving spin-$1/2$ DM particles considered in this work and makes the corresponding couplings dimensionless.
For momentum transfer $q$ and nucleon mass $m_N$, the momentum factors in the NREFT operators are
expressed in terms of $q/m_N$. 
We also use the following dictionary between LZ's notation for the isospin parameters and \texttt{WimPyDD}'s:

\begin{equation}
    (c_i^0,c_i^1)_{\mathrm{WimPyDD}}
    =
    \begin{cases}
        (2c_{i,\mathrm{LZ}}^s,0), & \text{isoscalar},\\
        (0,2c_{i,\mathrm{LZ}}^v), & \text{isovector}.
    \end{cases}
    \label{eq:isospin_conversion}
\end{equation}

For a target isotope $T$, the differential recoil rate is
\begin{equation}
    \frac{\mathrm{d}R_T}{\mathrm{d}E_R}
    =
    \frac{\rho_\chi}{m_\chi}
    \int_{u_\chi>u_{\chi, \min}}\mathrm{d}^3u_\chi\,
    f_{\mathrm{lab}}(\mathbf{u_\chi})\,u_\chi\,
    \frac{\mathrm{d}\sigma_T}{\mathrm{d}E_R}~,
    \label{eq:dd_rate}
\end{equation}
with the total rate obtained by summing over the naturally
occurring xenon isotopes weighted by their abundances (see, e.g., Ref.~\cite{Raj:2024guv}).
For spin-dependent scatters \texttt{WimPyDD} automatically accounts for the sum over isotopes with unpaired nucleons with appropriate structure factors.
We adopt the Standard Halo Model parameters used by LZ: the local DM density $\rho_\chi=0.3~\mathrm{GeV/cm^{3}}$, Galactic escape speed $u_{\mathrm{esc}}=544~\mathrm{km/s}$, local Galactic rotational velocity $\mathbf{v}_{\mathrm{rot}}=(0,238,0)~\mathrm{km/s}$, and solar peculiar velocity $\mathbf{v}_{\odot}=(11.1,12.2,7.3)~\mathrm{km/s}$. We evaluate the annually averaged velocity integral using the Maxwell–Boltzmann distribution implemented in \texttt{WimPyDD}.

To reproduce the nuclear input of the LZ EFT analysis, we replace the
default \texttt{WimPyDD} xenon responses by responses constructed from
the updated GCN5082 one-body density matrices used by
LZ~\cite{LZ:2023eft}. 
The density matrices for the relevant xenon
isotopes are obtained from the public \texttt{Elastic} repository of
the Berkeley Electroweak Physics group~\cite{BerkeleyElastic}.
We use
\texttt{dmscatter}~\cite{Gorton:2022dsc} that is based on \texttt{DMFormfactor}~\cite{Anand:2013yka} to generate the corresponding
momentum-dependent nuclear response functions. 
These responses, based on full-basis GCN5082
shell-model calculations~\cite{PandaX-II:2018xnv}, are used for every operator we consider.
We then fold the predicted spectrum for operator $a$ in Table~\ref{tab:opbenchmarks} the nuclear recoil efficiencies $\epsilon_{\rm LZ}(E_R)$ provided by LZ in Ref.~\cite{LZ:2026axp},
\begin{equation}\label{eq:accepted_rate}
    \frac{\mathrm{d}R^{\rm acc}_{a}}{\mathrm{d}E_R} = \epsilon_{\mathrm{LZ}}(E_R)
    \frac{\mathrm{d}R_a}{\mathrm{d}E_R} \, ,
\end{equation}
to get the expected signal yield as
\begin{equation}\label{eq:dd_events}
    \mu_a(m_\chi,\widehat d_a) =
    \mathcal{E}_{\mathrm{LZ}}
    \int_{E_{\min}}^{E_{\max}}\mathrm{d}E_R\,
        \frac{\mathrm{d}R^{\rm acc}_a(E_R;m_\chi,\widehat d_a)}{\mathrm{d}E_R} \, ,
\end{equation}
where the exposure $\mathcal{E}_{\rm LZ} = 4.71~{\rm tonne~ (fiducial~mass)} \times$ 220~live~days, 
$E_{\rm min} = 5.4~$keV,
$E_{\rm max} = 269.9~\mathrm{keV}$~\cite{LZ:2026axp}. 

The full LZ analysis uses an unbinned profile likelihood in the
$\{S1c,\log_{10}(S2c)\}$ plane~\cite{LZ:2026axp}. 
The detector
simulation, nuisance parameter model, veto samples, and toy Monte Carlo
distributions required to reproduce that likelihood are not fully
public.
We therefore construct an approximate energy-dependent likelihood and calibrate its effective background density using the published LZ significance.
To do this, we take the central preferred value and the net uncertainty~\cite{LZ:2026axp},
\begin{equation}
    E_c=248~\mathrm{keV},
    \qquad
    \sigma_E=\sqrt{23^2+23^2}~\mathrm{keV}
    \simeq32.5~\mathrm{keV},
\end{equation}
and for a given operator at some $m_\chi$ we define an effective normalized signal density at the candidate energy as
\begin{equation}
    f_a(E_c;m_\chi)
    =
    \frac{
        \displaystyle
        \int_{E_{\min}}^{E_{\max}}\mathrm{d}E_R\,
        \frac{\mathrm{d}R_{a,\mathrm{acc}}}{\mathrm{d}E_R}
        \mathcal{G}(E_c,E_R;\sigma_E)
    }{
        \displaystyle
        \int_{E_{\min}}^{E_{\max}}\mathrm{d}E_R\,
        \frac{\mathrm{d}R_{a,\mathrm{acc}}}{\mathrm{d}E_R}
    },
    \label{eq:signal_pdf_candidate}
\end{equation}
where $\mathcal{G}$ is a unit-normalized Gaussian resolution kernel, giving $f_a$ units of inverse recoil energy. 

For LZ230616 we use the extended
likelihood
\begin{equation}\label{eq:one_event_likelihood}
    \ln\mathcal{L}_a(\mu_a) = -(\mu_a+B) + \ln\!\left[
    \sigma_E\bigl(\mu_a f_a(E_c;m_\chi)+\beta_a\bigr) \right] \, .
\end{equation}
where $B$ is the number of background events and $\beta_a$ is an effective local background density in this one-dimensional projection. For $f_a>\beta_a$, maximizing Eq.~\eqref{eq:one_event_likelihood} gives
\begin{equation}\label{eq:muahat}
    \widehat\mu_a =1-\frac{\beta_a}{f_a} \, ,
\end{equation}
while for
$f_a\leq\beta_a$,  $\widehat\mu_a=0$. 
We then define
\begin{equation}
    q_a(\mu_a) \equiv -2\ln\!\left[
        \frac{\mathcal{L}_a(\mu_a)}
             {\mathcal{L}_a(\widehat\mu_a)}~ \right] \, .
    \label{eq:dd_test_statistic}
\end{equation}
\begin{equation}
    m_{\chi}^{\rm ref} = 1000~\mathrm{GeV}.
\end{equation}
The effective background density $\beta_a$ is chosen so that the background-only test statistic reproduces the corresponding operator's local significance $Z_a^{\mathrm{LZ}}$ reported at
$m_{\mathrm{ref}}$ in Table~S7 of Ref.~\cite{LZ:2026axp}:
\begin{equation}\label{eq:q0_significance}
    q_{0,a} \equiv -2\ln\!\left[
        \frac{\mathcal{L}_a(0)}
             {\mathcal{L}_a(\widehat\mu_a)}
    \right] = \left(Z_a^{\mathrm{LZ}}\right)^2 \, .
\end{equation}
Using Eqs.~\eqref{eq:one_event_likelihood} and \eqref{eq:muahat},
\begin{eqnarray}
 \nonumber   \left(Z_a^{\mathrm{LZ}}\right)^2
    &=&
    2\left[
        \ln\!\left(\frac{f_{a}^{\rm ref}}{\beta_a}\right)
        -1+
        \frac{\beta_a}{f_{a}^{\rm ref}}
    \right],\\
        f_{a}^{\rm ref}
    &\equiv& f_a(E_c;m_{\chi}^{\rm ref}).
    \label{eq:background_calibration}
\end{eqnarray}
Equivalently, defining
$r_a\equiv\beta_a/f_{a}^{\rm ref}$ with $0<r_a<1$, we solve
\begin{equation}
    2\left[-\ln r_a-1+r_a\right]
    =\left(Z_a^{\mathrm{LZ}}\right)^2,
    \label{eq:background_ratio_calibration}
\end{equation}
which makes explicit that significances published by LZ fixes the {\em ratio of
 the effective local background density to the signal density}. 
That is, the calculated recoil spectra themselves are not rescaled to match some LZ spectrum.

At each $m_\chi$, the best-fit coupling is obtained from
\begin{equation}
    \widehat d_a^{\,\mathrm{best}}
    =
    \sqrt{
        \frac{\widehat\mu_a}
             {\mu_a(m_\chi,1)}
    },
\end{equation}
with the lower and upper boundaries on either side of the best fit obtained from the two solutions of
\begin{equation}
    q_a\!\left[
        (\widehat d_a)^2\mu_a(m_\chi,1)
    \right]
    =2.71,
    \label{eq:dd_coupling_limits}
\end{equation}
as appropriate for an $90\%$~confidence threshold for one
parameter.
A non-zero lower boundary exists only
when $q_a(0)>2.71$; otherwise the background-only hypothesis lies
within the $90\%$ confidence interval and only an upper limit is
obtained. 

In Fig.~\ref{fig:L10_O1s_dd_limits_LZ} left panel we show our recoil spectra for the benchmark parameters used by LZ in Ref.~\cite{LZ:2026axp}, as well as LZ's evaluation of the same; these are seen to be in excellent agreement.
In the right panel of this figure we then show our best-fit parameters and 90\% confidence intervals, as reconstructed using our recipe above, for the elastic operator $\mathcal{L}_{10}^s$ and inelastic operator $\mathcal{O}_1^s$, superimposing LZ's curves from Ref.~\cite{LZ:2026axp}.
Once again we find good agreement with deviations by a factor of only a few.
In Appendix~\ref{app:dd_compare_all_ops} we provide a comparison of our reconstructed confidence regions for all the elastic operators listed in Table~\ref{tab:opbenchmarks} with the numerical contours in the official LZ DataRelease~\cite{LZ:HEPData2026} that accompanies Ref.~\cite{LZ:2026axp}.

%============== END: Direct detection ================%

%
%
%=================== Fig: 3 =====================%
\begin{figure*}[t]
    \centering
    \begin{subfigure}{0.328\textwidth}
        \centering
        \includegraphics[width=0.975\linewidth]{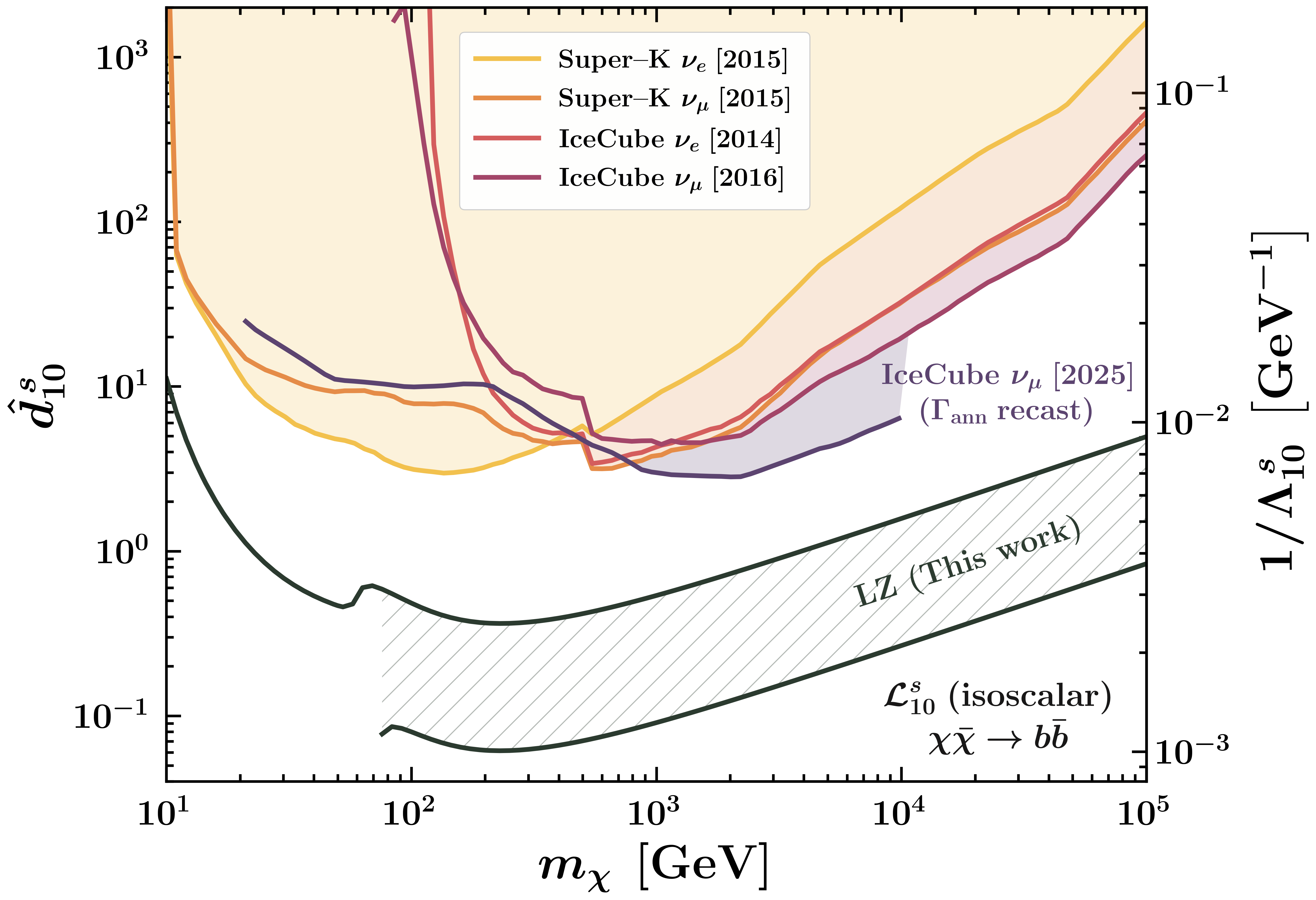}
        \label{sf:Ls10_capture_limits_bb}
        % \caption{}
    \end{subfigure}
    \begin{subfigure}{0.328\textwidth}
        \centering
        \includegraphics[width=0.975\linewidth]{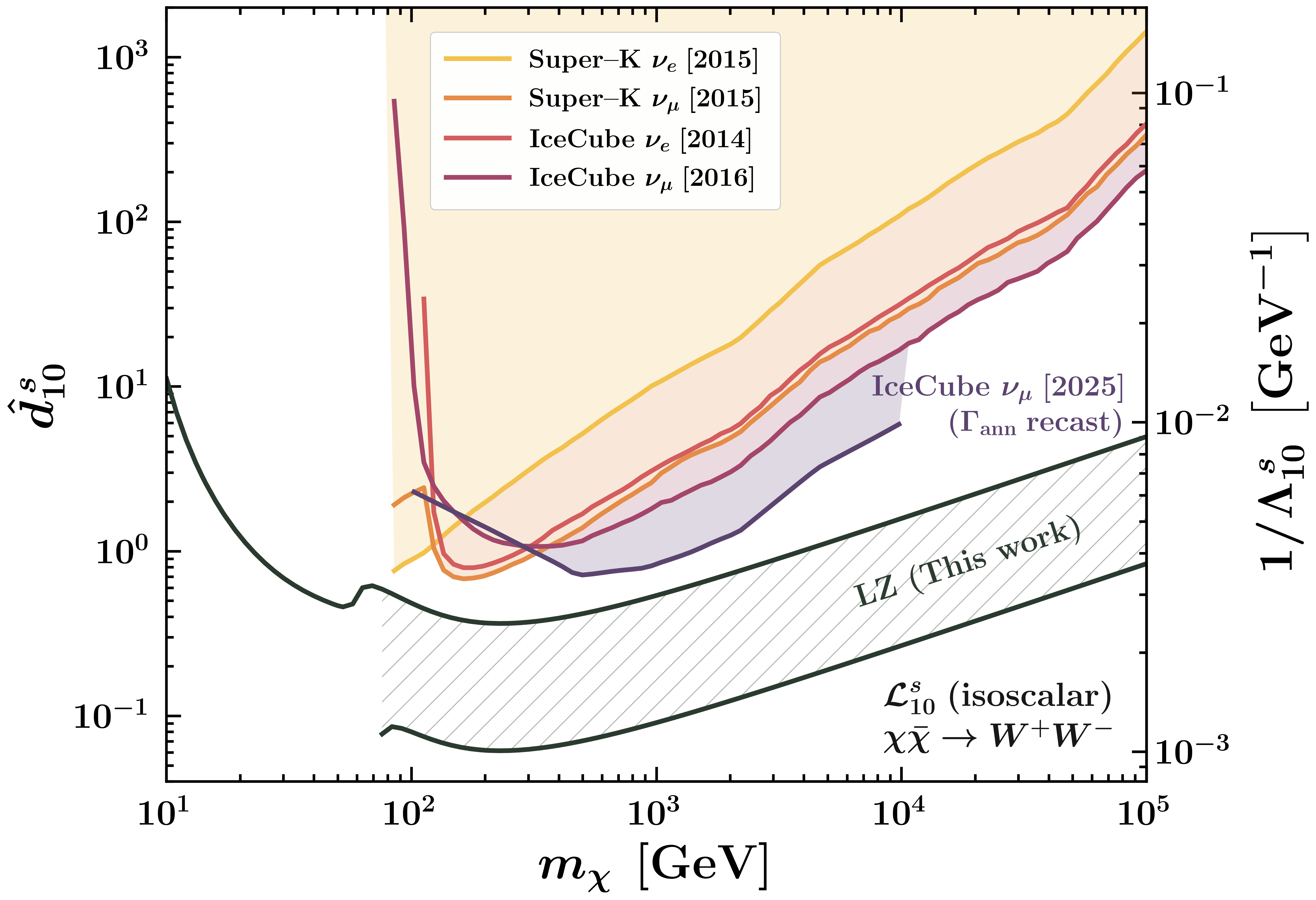}
        \label{sf:Ls10_capture_limits_WW}
        % \caption{}
    \end{subfigure}
    \begin{subfigure}{0.328\textwidth}
        \centering
        \includegraphics[width=0.975\linewidth]{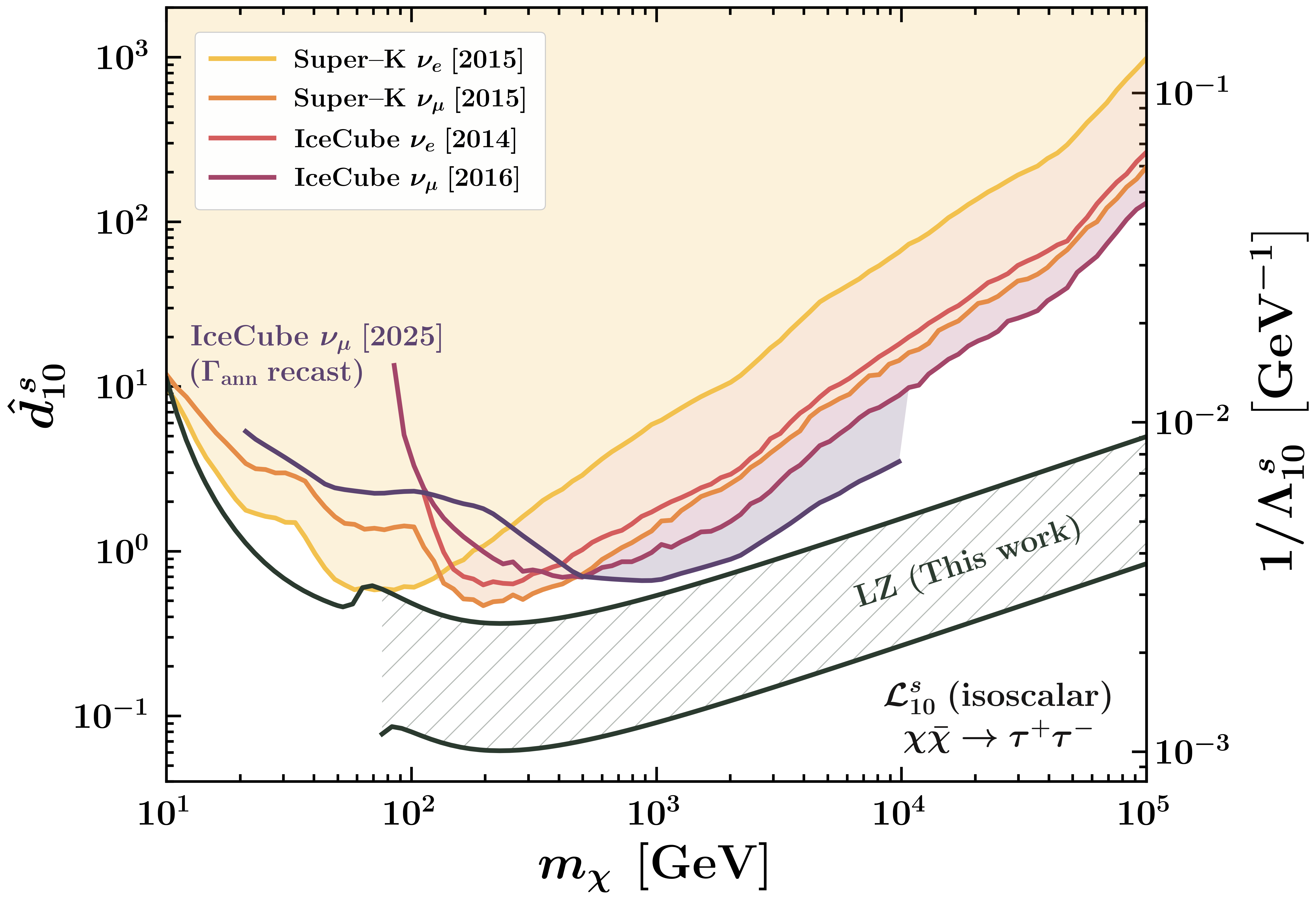}
        \label{sf:Ls10_capture_limits_tautau}
        % \caption{}
    \end{subfigure}
%
%%%%%
    \caption{\justifying 95\% C.L. limits from IceCube~\cite{IceCube:2015mgt,IceCube:2016dgk,IceCube:2025fcu} and Super-K~\cite{Super-Kamiokande:2015qek} measurements on the Wilson coefficient of the NREFT elastic scattering operator $\mathcal{L}_{10}^s$ as a function of DM mass from DM solar capture followed by annihilations to various final states that produce neutrino fluxes; see Fig.~\ref{fig:Ls10_capture_limits_nunu} for analogous limits on the direct $\nu\bar{\nu}$ channel.
    The 2025 solar-direction IceCube $\nu_\mu$ measurements are recast by rescaling the limits in Ref.~\cite{IceCube:2025fcu} as described in Sec.~\ref{sec:analysisresults}. 
    For comparison is shown the best-fit parameters and 90\% confidence interval of LZ230616 derived in Sec.~\ref{sec:lzlikelihoods} and displayed in Fig.~\ref{fig:L10_O1s_dd_limits_LZ}.
    The $b\bar{b}$ final state gives the weakest limits, and the $\nu\bar{\nu}$ final state the strongest limits, with the latter ruling out a good fraction of the LZ best-fit parameter space. 
    See Sec.~\ref{sec:analysisresults} for further details.}
    \label{fig:L10_capture_limits}
\end{figure*}
%================================================%
%
%
%=================== Capture rate and annihilation ================%
\section{Dark matter capture and annihilation in the Sun}
\label{sec:captureannihilations}

The DM--nucleon scatters sought in direct detection setups can also be probed in celestial bodies.
Halo DM states falling into the gravitational potential of the Sun can scatter on solar constituents and lose sufficient kinetic energy to become gravitationally bound.
We estimate rate of this capture using \texttt{WimPyC}~\cite{Kang:2025yci}, an extension of \texttt{WimPyDD}.
It allows for the direct use of the NREFT Hamiltonian with the same couplings, normalization, and isospin convention adopted in Sec.~\ref{sec:lzlikelihoods}. 
The DM solar capture rate is given by~\cite{Kang:2025yci}
\begin{equation}\label{eq:cap_rate}
\begin{aligned}
    C^\odot_\chi = \sum_T \dfrac{\rho_\chi}{m_\chi} \int_0^{R_\odot} 4\pi r^2 \, dr
    & \int_0^{u_{\chi, \max}} du_\chi \, \dfrac{f_\odot(u_\chi)}{u_\chi} \, w(u_\chi, r)^2 \\
    &  \times n_T(r) \int_{E_R^{\min}}^{E_R^{\max}} dE_R\, \frac{d\sigma_{\chi T}}{dE_R} \,,
\end{aligned}
\end{equation}
where $R_\odot$ is the solar radius, $n_T(r)$ is the number density of solar isotope $T$ at radius $r$,
$u_\chi$ ihe asymptotic DM speed far from the Sun and 
$v_{\rm esc}(r)$ is local solar escape speed, with $w^2(u_\chi,r)=u_\chi^2+v_{{\rm esc}, \odot}^2(r)$ giving the DM speed at radius $r$;
$f_\odot(u_\chi)$ is the normalized DM speed distribution in the solar rest frame, obtained from a Maxwell--Boltzmann distribution in the Galactic rest frame (see Ref.~\cite{Bose:2022ola} for the effects of deviations from this assumption). 
The maximum incident speed $u_{\chi,\max}$ is fixed by the requirement that capture be kinematically possible.
The recoil energy limits $E_R^{\min}$ and $E_R^{\max}$ are given in Eqs.~(3)--(6) of Ref.~\cite{Kang:2025yci}. 

We evaluate the differential cross section in Eq.~\eqref{eq:cap_rate} using the same NREFT Hamiltonian employed in the LZ analysis for elastic scattering, and the same halo parameters as in Sec.~\ref{sec:lzlikelihoods}. 
For the solar structure, we use the predefined \texttt{WimPyC} Sun model, which implements the radial density and elemental abundance profiles of the AGSS09 solar model~\cite{Serenelli:2009yc}.
For operators containing only spin-dependent nuclear responses, we only include solar isotopes with non-zero nuclear spin: $^1{\rm H}$, $^3{\rm He}$, $^{13}{\rm C}$, $^{14}{\rm N}$, $^{15}{\rm N}$, $^{17}{\rm O}$, $^{23}{\rm Na}$, and $^{27}{\rm Al}$, and for operators involving spin-independent scattering we use the complete solar isotope set in \texttt{WimPyC}.
For each interaction, we calculate $C_\chi^\odot$ at the reference coupling $\widehat d_a=1$ over the mass range $10~{\rm GeV}\leq m_\chi\leq10^5~{\rm GeV}$ and then scale the capture rate quadratically with the coupling, which is expected for the range of couplings we constrain.
Numerical interpolation introduces some jaggedness in the capture rate as a function of $m_\chi$, due to which we use three mass points per decade to reduce its impact while accounting for the overall decrease of $C_\chi^\odot$ with increasing $m_\chi$.
Some residual unevenness remains and is reflected in the limits we display.

Captured $\chi$s can undergo further scatters with solar constituents, lose energy, and eventually thermalize in the solar core. 
Neglecting evaporation, as appropriate for the DM masses considered here~\cite{Garani:2021feo}, the number of captured $\chi$s evolves as
\begin{equation}\label{eq:dNdt}
    \frac{dN_\chi}{dt} = C_\chi^\odot-C_{\rm ann}N_\chi^2 \, ,
\end{equation}
where $C_{\rm ann}=\langle\sigma v\rangle/V_{\rm eff}$ is the annihilation coefficient, with $V_{\rm eff}$ the effective volume occupied by the thermalized DM population. 
For the various NREFT operators $\mathcal{O}_i$~\cite{Catena:2015iea, Catena:2015uha, Catena:2015vpa, Catena:2016ckl}, thermalization of the DM with the solar core and capture-annihilation equilibrium is discussed in Refs.\,\cite{AvisKozar:2023iyb, GAMBIT:2021rlp, Widmark:2017yvd}. 
As the elastic operators in Table~\ref{tab:opbenchmarks} give rise, in the non-relativistic limit, to combinations of $\mathcal{O}_i$s weighted by momentum factors, we made use of Ref.\,\cite{Widmark:2017yvd} to check the corresponding thermalization times. 
For the Wilson coefficients that we exclude, we find that DM achieves thermalization within the age of the Sun for all the operators we consider.
The annihilation rate then follows from the solution of Eq.~\eqref{eq:dNdt} as
\begin{equation}\label{eq:gam_ann}
    \Gamma_{\rm ann} = \frac{1}{2}C_{\rm ann}N_\chi^2(t_\odot) = \frac{1}{2}C_\chi^\odot \tanh^2 \left(\frac{t_\odot}{\tau_{\rm eq}}\right)\, ,
\end{equation}
where $t_\odot$ is the age of the Sun, the equilibrium timescale $\tau_{\rm eq} = 1/\sqrt{C_\chi^\odot C_{\rm ann}}$ and the factor $1/2$ accounts for the removal of two particles ($\chi$ and $\bar{\chi}$) in each annihilation.
We use the annihilation rate, $\Gamma_{\rm ann}$ defined in Eq.~\eqref{eq:gam_ann} throughout our analysis.
However, over most of the parameter space constrained through solar capture in this work, capture--annihilation equilibrium is reached within the solar age, yielding $\Gamma_{\rm ann} = C_\chi^\odot/2$. 
Outside this regime, the factor $\tanh^2(t_\odot/\tau_{\rm eq})$ consistently accounts for the suppression of the neutrino flux.
We consider one annihilation channel at a time, $\chi \bar{\chi} \rightarrow f$, with $f\in\{W^+W^-,\,b\bar b,\,\tau^+\tau^-,\,\nu\bar\nu\}$. The differential neutrino flux at Earth is
\begin{equation}
    \frac{d\Phi_{\nu_\alpha+\bar\nu_\alpha}}{dE_\nu} = \frac{\Gamma_{\rm ann}}{4\pi ({\rm AU})^2}
    \left[ \frac{dN_{\nu_\alpha+\bar\nu_\alpha}^{\,f}} {dE_\nu} \right]_{\oplus} \, ,
\end{equation}
where the quantity in square brackets denotes the neutrino yield per annihilation for the final state $f$, after propagation from the solar core to Earth. When deriving the limit for a given channel, we assume a branching fraction of unity into that channel. For some UV-complete model with different branching ratios to various SM final states, our limits can be scaled accordingly.

We generate the neutrino spectra using $\chi\texttt{aro}\nu$~\cite{Liu:2020ckq}, which accounts for the decay, hadronization, and interactions of the annihilation products in the solar medium, as well as electroweak corrections. The neutrinos are subsequently propagated from the solar core to Earth using \texttt{nuSQuIDS}~\cite{Arguelles:2021twb}, including flavor oscillations, charged and neutral current interactions, and neutrino regeneration through $tau$-lepton production and decay. We retain the neutrino and antineutrino spectra separately during propagation and combine them when estimating the detector-level signal. 
The resulting
$\nu_\mu+\bar\nu_\mu$ spectra at Earth are used below to evaluate the expected Super-K and IceCube event distributions.
%
%============== END: Capture rate and annihilation ================%

%
%
%=================== Fig: 4 =====================%
\begin{figure*}[t]
    \centering
    \includegraphics[width=1.0\linewidth]{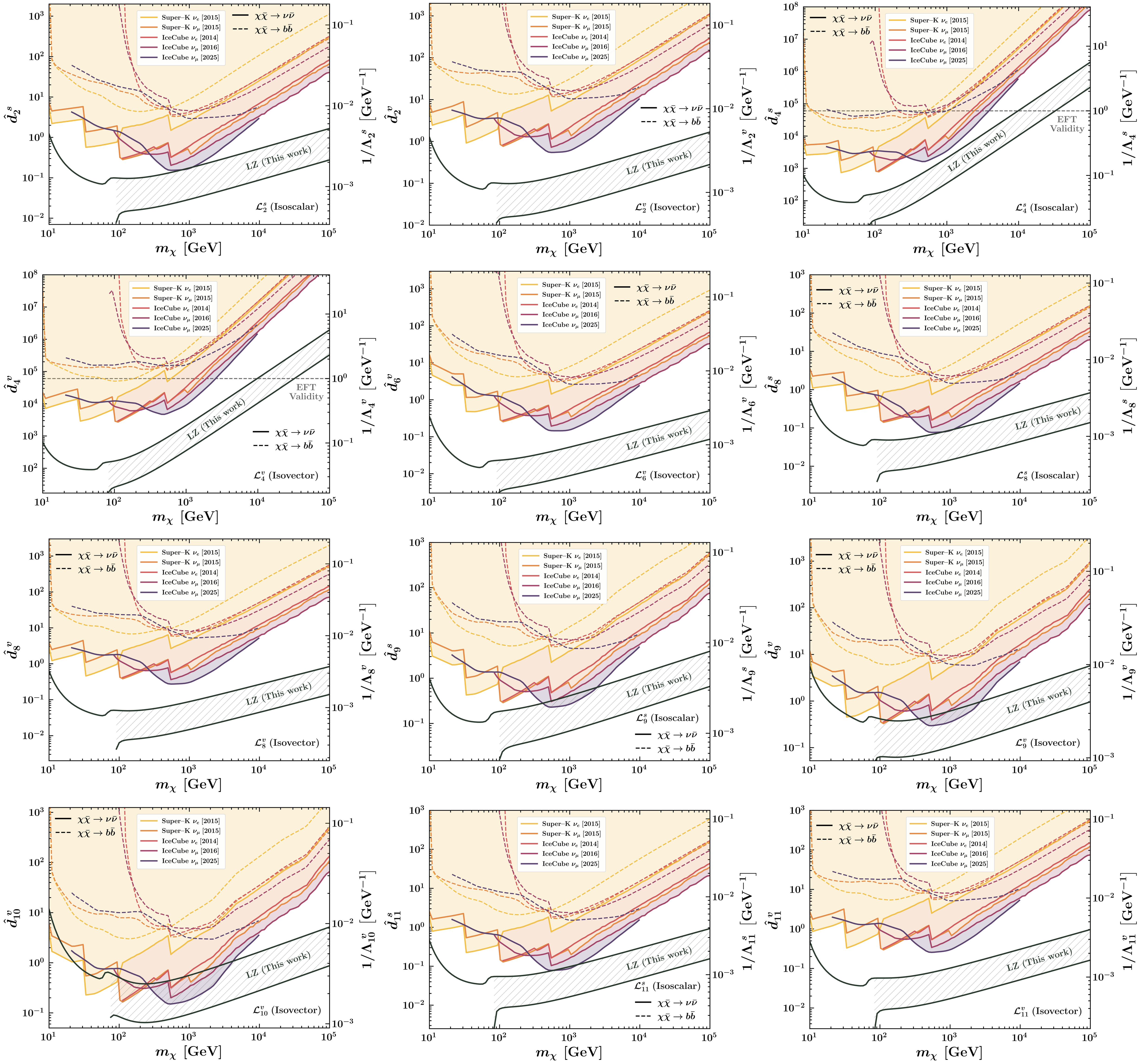}
    \caption{\justifying Same as Figure~\ref{fig:L10_O1s_dd_limits_LZ}, but for the rest of the operators listed in Table~\ref{tab:opbenchmarks}, and with only the final states that give the strongest ($\nu\bar{\nu}$) and weakest ($b\bar{b}$) limits.
    This figure is continued in Fig.~\ref{fig:all_ops_limits_combo_2}.
    See Sec.~\ref{sec:analysisresults} for further details. 
    For most operators, the EFT prescription is valid throughout the displayed parameter range. For the others, it is valid only below the horizontal dashed line, corresponding to the cutoff exceeding GeV momentum transfers in scattering.}
    \label{fig:all_ops_limits_combo_1}
\end{figure*}
%================================================%
%
%
%
%
%=================== Fig: 5 =====================%
\begin{figure*}[t]
    \centering
    \includegraphics[width=1.0\linewidth]{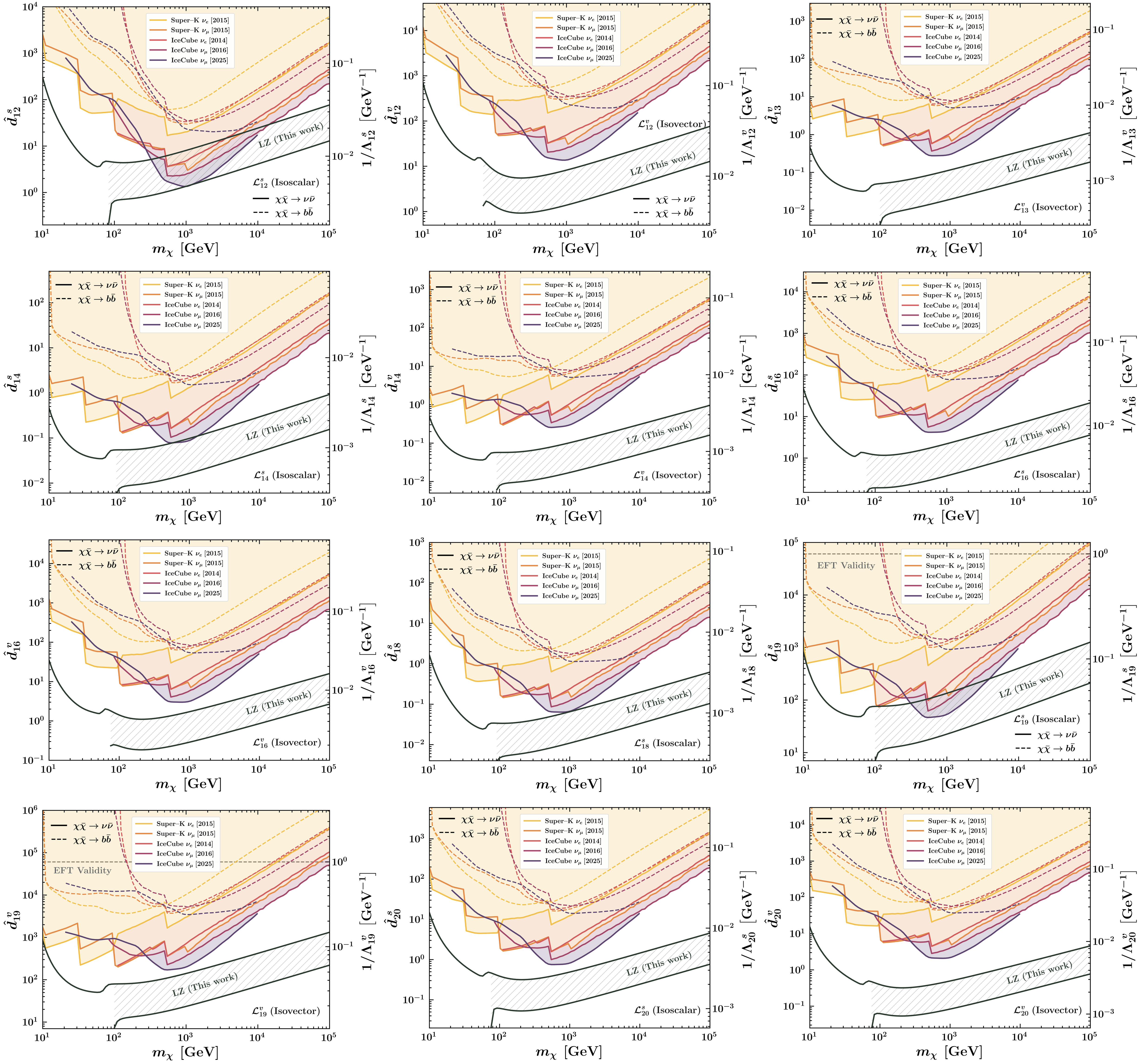}
    \caption{\justifying Continuation of Fig.~\ref{fig:all_ops_limits_combo_1}. 
    See Sec.~\ref{sec:analysisresults} for further details.}
    \label{fig:all_ops_limits_combo_2}
\end{figure*}
%================================================%
%
%

%================= Analysis ==================%
\section{Analysis \& results}
\label{sec:analysisresults}

To set our limits we use high energy neutrino datasets from IceCube and Super-Kamiokande, which were also used in Ref.\,\cite{Bose:2026ndd}. 
For completeness, here we briefly describe them.

\textbf{(i) Atmospheric $\nu_\mu$ and $\nu_e$ flux (Super-K\,-\,2015)}~\cite{Super-Kamiokande:2015qek}:
 We use the atmospheric $\nu_e$ and $\nu_\mu$ fluxes measured in Runs I-IV, provided in Ref.\,\cite{Super-Kamiokande:2015qek}, Figure 7, and the HKKM11 atmospheric background model\,\cite{Honda:2006qj, PhysRevD.83.123001}. 
 To subtract overwhelming backgrounds from non-solar directions we use an angular cut: 
 we multiply the total all-sky per-steradian flux by 
\begin{eqnarray}\label{eq: angular cut}
    2\pi \int_{0}^{\delta\theta(E_\nu)}
    d\theta\,\sin\theta
    &=&
    4\pi \sin^2\left(
    \frac{\delta\theta(E_\nu)}{2}
    \right) \, ,
\end{eqnarray}
where $\delta\theta(E_\nu)$ is the energy-dependent angular resolution at Super-K.
See Ref.~\cite{Krishna:2025ncv} for more details on the angular cut. 

\textbf{(ii) Atmospheric $\nu_e$ flux (IceCube\,-\,2014)}~\cite{IceCube:2015mgt}:
We use the dedicated atmospheric $\nu_e$ flux measurements and the ``modified Honda'' background model in Ref.\,\cite{IceCube:2015mgt}. 
As with the Super-K analysis, here too we use an angular cut to focus on the solar-direction neutrinos.

\textbf{(iii) Dedicated directional $\nu_\mu$ flux (IceCube\,-\,2016)}~\cite{IceCube:2016dgk}:
Here IceCube focuses on high energy upgoing $\nu_\mu$s from the solar direction, for which IceCube took dedicated measurements of up-going track events over a total livetime of $T = 532 \ {\rm days}$ when the Sun was below the horizon.
The flux measurements are given in Ref.\,\cite{IceCube:2016dgk} as a function of the solar opening angle $\theta_\odot$. 
We closely follow Ref.\,\cite{IceCube:2016dgk} to evaluate the DM-originated signal flux across angular bins.
The differential event rate is
\begin{equation}\label{eq:event_rate_theta}
\begin{aligned}
    \frac{dN_{\theta_\odot}}{d\cos \theta_\odot}
    &=
    2T
    \int_{E_\nu^{\min}}^{E_\nu^{\max}}
    A_{\rm eff}(E_\nu)
    \frac{d\Phi_{\nu +\bar{\nu}}}{dE_\nu}
    \frac{1}{\sqrt{2\pi}\sigma_\theta}
    \\
    &\qquad\times
    \exp\left[
    -\frac{\left(1-\cos \theta_\odot \right)^2}
    {2\sigma_\theta^2}
    \right]
    \, dE_\nu \,,
\end{aligned}
\end{equation}
where $A_{\rm eff} (E)$ is the effective area~\cite{IceCube:2016dgk}, and the energy-dependent angular dispersion is
\begin{equation}
    \sigma_\theta
    =
    \left|
    \frac{
    \sqrt{2}\left(1-\cos\left[\Delta\theta(E_\nu)\right]\right)
    }{
    2\,{\rm erf^{-1}}(0.5)
    }
    \right| \,\,,
\end{equation}
where we use Ref.\,\cite{IceCube:2016dgk} for the the median energy resolution.  

\textbf{(iv) Dedicated directional $\nu_\mu$ flux (IceCube\,-\,2025)}~\cite{IceCube:2025fcu}:
This analysis again uses dedicated observations of muon-like tracks from the direction of the Sun.
Due to lack of information for reproducing the analysis pipeline, we recast the spin-independent scattering limits in Ref.\,\cite{IceCube:2025fcu} by scaling them according to the capture rate from Ref.\,\cite{Garani:2017jcj}.
This results in an upper limit on the annihilation rate, related to our capture rates through Eq.~\eqref{eq:gam_ann}, which we then use to set limits on our operators.
\\

For each of the datasets above we calculate 

\begin{equation}\label{eq:chisq}
    \chi^2 = \sum_{i=1}^{N_{\rm bin}} \frac{\left(\Phi_{\rm d}^i - \Phi_{\rm atm}^i - \Phi_{\chi}^i\right)^2}{{\sigma_{\rm d}^i}^2} \, ,  
\end{equation}
where  $\Phi_{\rm d}^i$ is the measured $\nu + {\bar \nu}$ flux, $\sigma_{\rm d}^{i}$ is the corresponding uncertainty, $\Phi_{\rm atm}^i$ is the atmospheric neutrino model flux (with the corresponding angular cuts), and $\Phi_{\chi}^i$ is our signal neutrino flux at the $i^{\rm th}$ bin. We obtain the 95\% C.L. upper limits on the corresponding couplings of the operators of interest by setting  $\chi^2-\chi_{\rm min}^2 = 2.71$. Here $\chi_{\rm min}^2$ is the minimum obtained by varying our model parameters~\cite{Lamperstorfer:2015cfg}.
For IceCube-2016 dataset, instead of the flux we use event counts. 
We note here that DeepCore\,-\,2016\,\cite{IceCube:2016dgk} and DeepCore\,-\,2021\,\cite{IceCube:2021xzo} datasets do not probe parameter space that is not already covered by our analysis, owing to which we do not show these limits.

In Figs.~\ref{fig:Ls10_capture_limits_nunu} and \ref{fig:L10_capture_limits} we show for the operator $\mathcal{L}_{10}^s$ our 95\% C.L. limits for all DM annihilation final states studied by IceCube, superimposing the best-fit parameters that we derived in Sec.~\ref{sec:lzlikelihoods} and displayed in Fig.~\ref{fig:L10_O1s_dd_limits_LZ}.
The $\nu{\bar \nu}$ final state is seen to give the strongest limits and the $b\bar{b}$ final state the weakest limits; the hierarchy of limits of various final states follows from the neutrino fluxes they respectively produce.
On the right $y$-axis we show values of the EFT cutoff $\Lambda$ corresponding to the Wilson co-efficients on the left $y$-axis, obtained by using the dictionary in Eq.~\eqref{eq:lagrangian_normalization}, which gives $\Lambda = v_{\rm H}/\sqrt{\hat{d}}$. 
With dashed horizontal lines we also show the region in which the EFT description is valid, set by the maximum momentum transfer $q_{\rm max}$ in DM-nucleus scatters being smaller than the $\Lambda$.
For scattering on heavy nuclei in the solar core we expect $q_{\rm max}$ up to about GeV, and accordingly we show the region of the validity of the EFT to be where $\Lambda<$1~GeV.
Apart from $\mathcal{L}_4$, the LZ-probed regions for all operators lie within the EFT-valid regime. For $\mathcal{L}_{19}$, a very small portion of the capture-induced bounds extends beyond the EFT-valid regime.

In Figs.~\ref{fig:all_ops_limits_combo_1} and \ref{fig:all_ops_limits_combo_2} we then show our limits for the rest of the operators in Table~\ref{tab:opbenchmarks}, restricting ourselves to only $\nu\bar{\nu}$ and $b\bar{b}$ final states as they provide the strongest and weakest bounds respectively.
For several of our operators we see that the IceCube and Super-K limits disfavor large ranges of parameters that fit the LZ230616 event.
%
%============== END: Analysis ================%

%================= Discussion ==================%
\section{Discussion}
\label{sec:discs}

In this work we have shown that solar capture is a potential complementary probe by which effective operators that may fit the LZ230616 event with high significance may be tested.
While we have assumed standard halo velocities (see Ref.~\cite{OHare:2026nqi} for a simulated study of the maximal DM speeds on the day of the LZ event), DM in the solar vicinity may receive boosts from the motion of the Large Magellanic Cloud~\cite{Fan:2026kxx} or unvirialized substructure~\cite{OHare:2014nxd,OHare:2018trr,Maity:2022enp, Aggarwal:2024ngx}, which would impact both direct detection (hence the significance of the fits of velocity/momentum-dependent operators) and solar capture.
Further, while the NREFT of Refs.~\cite{Fitzpatrick:2012ix,Anand:2013yka,Barello:2014uda} was constructed explicitly for direct detection analyses at the nucleon level, one could also envisage fitting the LZ event with {\em quark-level} effective operators.
(Anti-)quarks would be the natural degrees of freedom as final states of DM self-annihilations in the Sun occurring at energies higher than the QCD scale, and due to their hadronization a neutrino flux would automatically arise without having to specify new model parameters that result in desired final states.
The requirement of DM self-annihilations, on which rests our work and Ref.~\cite{Nguyen:2026lui}, assumes a particle-antiparticle symmetric population of DM; the hypothesis that LZ230616 was caused by DM from an asymmetric population could be tested by such other means as the destruction of neutron stars~\cite{LZasymm}.

We leave these and other instructive questions to future study.

%
%============== END: Discussion ================%

%================== ACKNOWLEDGE ==================%
\section*{Acknowledgments}

We gratefully acknowledge Tim Linden, Dan Hooper, Biprajit Mondal 
and Rohan Pramanick for helpful discussions.
We thank B.~Ananthanarayan for unwittingly suggesting the \href{https://www.youtube.com/watch?v=QkF3oxziUI4}{title} of this work.
D.B. acknowledges the Anusandhan National Research Foundation (ANRF), Government of India, for supporting his research through the National Post-Doctoral Fellowship (N-PDF), File No.\,\,PDF/2026/003067, and also acknowledges support from the Council of Scientific and Industrial Research (CSIR), Government of India, under the Research Associateship program through grant no.\,\,09/0079(24106)/2025-EMR-I. 
T.T.Q.N. is supported by the Swedish Research Council under contract 2022-04283, and from EDUCATE Excellence Centre funded by the Swedish Research Council through grant Dnr~2022-06627. 
A.K.S. is supported by the NSF-BSF (grant no.\,\,2024829).
N.R. acknowledges support from the grant ANRF/ECRG/2024/000387/PMS and the Infosys Foundation, Bangalore.
R.L.\,\,acknowledges financial support from the institute start-up funds and ANRF for the grant no.\,\,ANRF/ARG/2025/005140/PS. 
%
%=============== END: ACKNOWLEDGE ================%

%=============== Appendix - DD Compare ================%
\appendix
%
%
%=================== Fig: 6 =====================%
\begin{figure*}[t]
    \centering
    \includegraphics[width=1.0\linewidth]{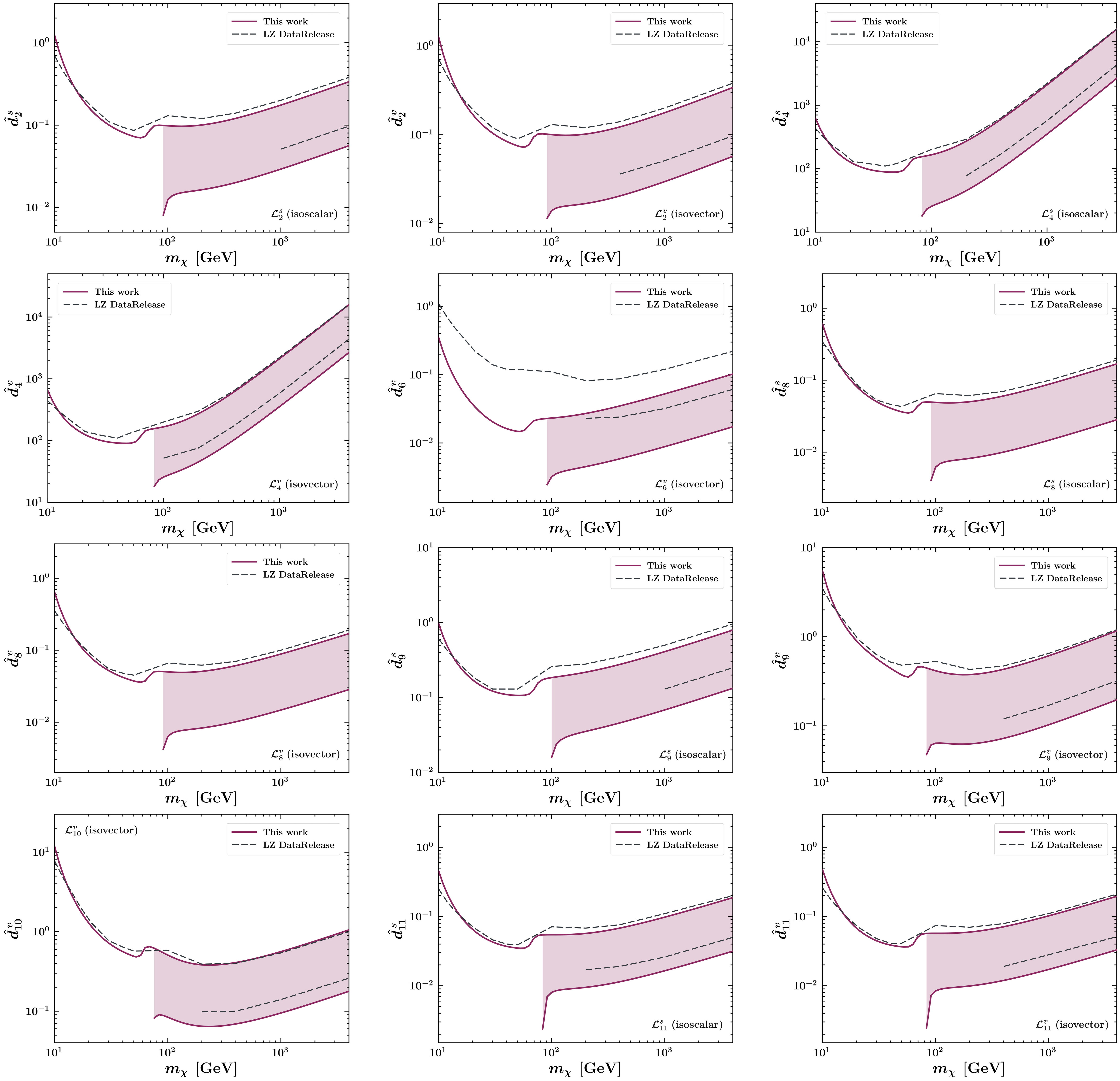}
    \caption{\justifying Comparison of the $90\%$ C.L. regions obtained with our reconstructed likelihood (solid magenta boundaries and shaded regions) with the numerical contours provided in the official LZ DataRelease~\cite{LZ:HEPData2026} accompanying Ref.~\cite{LZ:2026axp} (dashed gray curves), for the elastic operators listed in Table~\ref{tab:opbenchmarks}. This figure is continued in Fig.~\ref{fig:all_ops_limits_combo_dd_2}. See Appendix~\ref{app:dd_compare_all_ops} for further details.}
    \label{fig:all_ops_limits_combo_dd_1}
\end{figure*}
%================================================%
%
%
%
%
%=================== Fig: 7 =====================%
\begin{figure*}[t]
    \centering
    \includegraphics[width=1.0\linewidth]{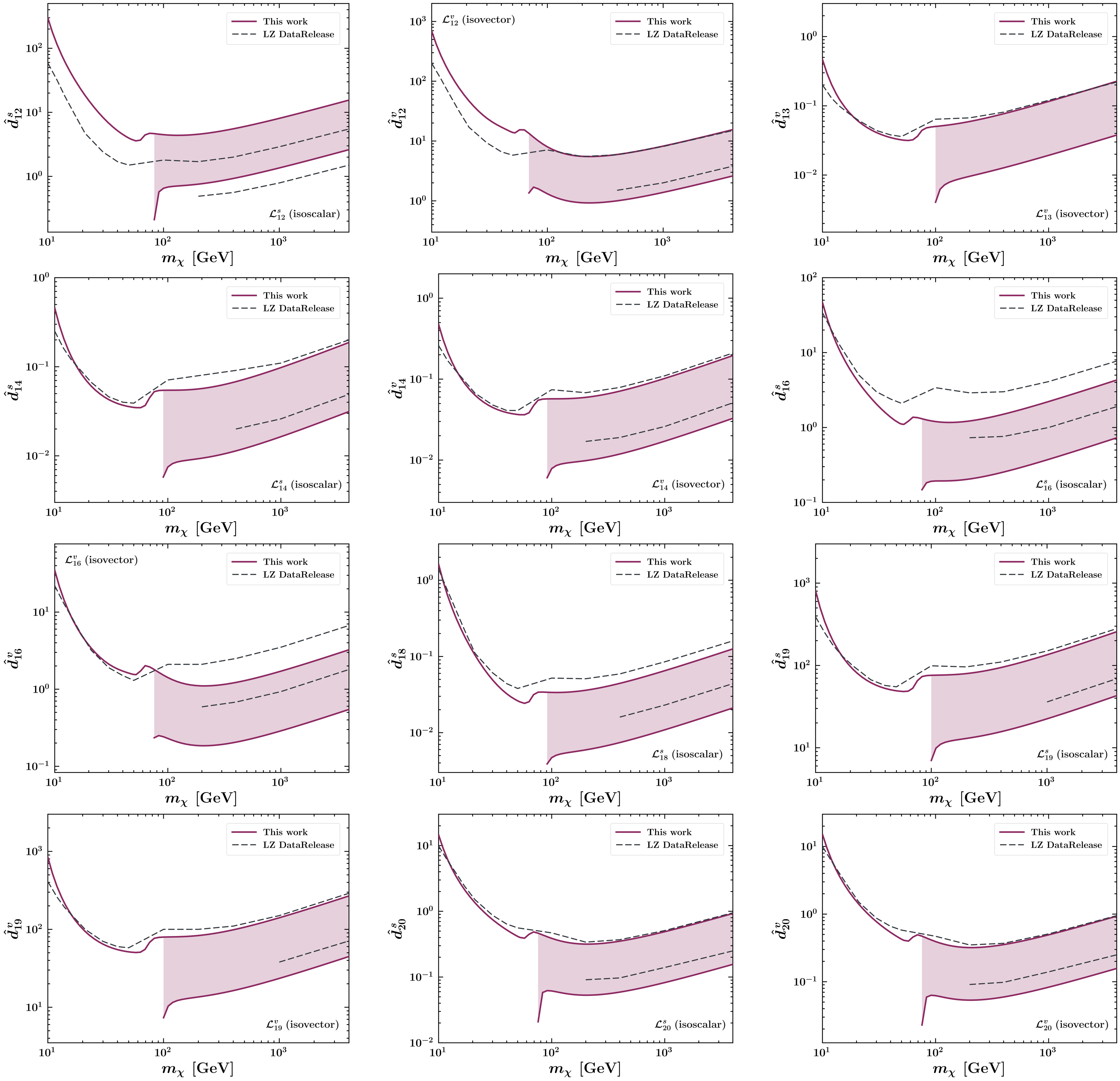}
    \caption{\justifying Continuation of Fig.~\ref{fig:all_ops_limits_combo_dd_1}. See Appendix~\ref{app:dd_compare_all_ops} for further details.}
    \label{fig:all_ops_limits_combo_dd_2}
\end{figure*}
%================================================%
%
%
\section{Theorists' reconstruction of LZ fits}
\label{app:dd_compare_all_ops}

The official
LZ DataRelease~\cite{LZ:HEPData2026} accompanying Ref.~\cite{LZ:2026axp} provides contours of
confidence intervals for the elastic operators considered in this work.
In Figs.~\ref{fig:all_ops_limits_combo_dd_1} and
\ref{fig:all_ops_limits_combo_dd_2}, we compare these contours with
the regions reconstructed using our likelihood method described in
Sec.~\ref{sec:lzlikelihoods}.
The regions for the $\mathcal{L}_{6}$, $\mathcal{L}_{12}$, and
$\mathcal{L}_{16}$ operators do not reproduce the LZ contours at the same level as the $\mathcal{L}_{10}^{s}$ result shown in
Fig.~\ref{fig:L10_O1s_dd_limits_LZ}. 
The largest difference occurs for
$\mathcal{L}_{6}^{v}$ at $m_\chi=200~\mathrm{GeV}$, where the ratios of the coupling between our reconstruction and LZ's are $5$ and $3$ for the lower and upper boundaries, respectively.  However, this
factor of few agreement is reasonable given that our reconstruction uses the recoil energy alone. 
A more precise reproduction would require the full numerical detector simulation and likelihood analysis in the
$\{S1c,\log_{10}(S2c)\}$ plane.
We remark here that for both $\mathcal{L}_{8}^{s}$ and $\mathcal{L}_{8}^{v}$, the LZ analysis quotes local significances of at least $3\sigma$ at the relevant benchmark masses, while the DataRelease tabulates only a nonzero upper contour.
%
%============= END: Appendix - DD Compare =============%

%================== References ===================%
\bibliography{refs}
%================= END: Refs =====================%

\end{document}